\documentclass{ws-procs9x6}

\begin{document}

\title{Vacuum condensates, 
effective gluon mass  
and color confinement  
}

\author{K.-I. Kondo\footnote{\uppercase{W}ork supported by \uppercase{S}umitomo Foundations, 
\uppercase{G}rant-in-Aid for \uppercase{S}cientific \uppercase{R}esearch (C)14540243 from \uppercase{J}apan \uppercase{S}ociety for the \uppercase{P}romotion of \uppercase{S}cience (\uppercase{JSPS}), 
and in part by \uppercase{G}rant-in-Aid for \uppercase{S}cientific \uppercase{R}esearch on \uppercase{P}riority \uppercase{A}reas (B)13135203 from
The \uppercase{M}inistry of \uppercase{E}ducation, \uppercase{C}ulture, \uppercase{S}ports, \uppercase{S}cience and \uppercase{T}echnology (\uppercase{MEXT}).}}

\address{Department of Physics, Faculty of Science, 
Chiba University, \\ Chiba 263-8522, Japan\\ 
E-mail: kondok@faculty.chiba-u.jp}


\maketitle

\abstracts{
We propose a new reformulation of Yang-Mills theory in which three- and four-gluon self-interactions are eliminated at the price of introducing a sufficient number of auxiliary fields.   
We discuss the validity of this reformulation in the possible applications such as dynamical gluon mass generation, color confinement and glueball mass calculation. 
We emphasize the transverse-gluon pair condensation as the basic mechanism for dynamical mass generation. 
 The confinement is realized as a consequence of a fact that the auxiliary fields become dynamical in the sense that they acquire the kinetic term due to quantum corrections. 
}

\section{Introduction}

In the classical level, Yang-Mills theory \cite{YM54} is a scale-invariant (actually, conformal invariant) field theory describing massless gauge boson. 
This is because the existence of the mass term is prohibited  in the classical Yang-Mills theory, once we require  {\it gauge invariance}.

  In the quantum level, on the other hand, this is not necessarily the case.  
To quantize the Yang-Mills theory, the procedure of gauge fixing (GF) is necessary.  The gauge-fixed Yang-Mills theory does not prohibit the mass term, since the gauge invariance no longer hold. Moreover, we need to introduce the Faddeev--Popov (FP) ghost field in the non-Abelian gauge group.  Consequently, the BRST invariance of the total Yang-Mills action is expected to hold in the quantum Yang-Mills theory instead of the gauge invariance.  
In fact, it is known \cite{CF76} that a mass term of the gauge boson  can be introduced in the Lagrangian by adding simultaneously the (gauge-parameter dependent) mass term of the ghost.  
In this case, the BRST transformation should be understood as the {\it on-shell} BRST transformation.  Here on-shell means that the Nakanishi-Lautrup field $B$ is eliminated by using the equation of motion for $B$. 
  However, the on-shell BRST transformation does not have the nilpotency, unless we use the equation of motion for the (anti)ghost.  
There is an option \cite{Ojima82} to modify the off-shell BRST transformation so that the mass term can be made off-shell invariant under the modified transformation by adopting an appropriate GF+FP term.  However, the off-shell nilpotency is lost in the modified off-shell BRST transformation. 
No one has found the mass term which is invariant under the off-shell BRST transformation with nilpotency.  
Remember that the nilpotency  plays the key role in the proof of the unitarity of the physical S-matrices.\cite{KO79}  Therefore, the failure of nilpotency endangers maintaining the unitarity of the theory. In fact, it is shown that the physical space specified by the subsidiary condition $Q_B' |phys \rangle=0$ using the modified BRST charge $Q_B'$ contains the negative norm state.  
These facts prevent us from pursuing the massive Yang-Mills theory at the tree level, i.e., Yang-Mills theory with a mass term in the Lagrangian level.  

In view of these, we are urged to consider the Yang-Mills theory whose mass is purely generated through quantum effects.  Thus we are lead to examine the dynamical mass generation of gauge boson in Yang-Mills theory.
In fact, such a possibility and its implications were investigated long ago in the following papers: 
[Eichten and Feinberg, 1974],
[Smit, 1974],
[Tye, Tomboulis and Poggio, 1975],
[Fukuda and Kugo, 1978],
[Gusynin and Miransky, 1978],
[Cornwall and Papavassiliou, 1982],
%
by using non-perturbative methods:
Schwinger-Dyson equation, 
Bethe-Salpeter equation, 
Effective potential, 
Variational technique, 
Cooper equation, 
Bogoliubov transformation, etc.
\noindent
However, I have no time to explain the details of these works. 

Here we mention recent studies of the coupled SD equations for the gluon and ghost propagators (in the Landau gauge) 
initiated by 
Smekal, Hauck and Alkofer.\cite{SHA97,AS01}
\noindent
They revealed the surprising results:
Euclidean propagators exhibit {\it infrared (IR) asymptotic power-law behaviors characterized by the critical exponent $\kappa$} ($1/2<\kappa <1)$:

\leftline{$\cdot$ \it IR suppression of the gluon propagator}
\begin{align}
 D_T(Q^2) \cong {A \over Q^2}(Q^2)^{2\kappa} \downarrow 0 
\quad \text{as} \quad Q^2 \downarrow 0 .
\end{align}

\leftline{$\cdot$ \it IR enhancement of the ghost propagator}
\begin{equation}
 \Delta_{FP}(Q^2) \cong {B \over Q^2}(Q^2)^{-\kappa} \uparrow \infty  \quad \text{as} \quad Q^2 \downarrow 0 .
\end{equation}

These approximate solutions lead to  

\leftline{$\cdot$ \it Existence of IR fixed point}
\begin{align}
  \beta(g^2(\mu)) \uparrow 0 \quad \text{as} \quad  \mu \downarrow 0 .\label{IRFP}
\end{align}

\leftline{$\cdot$ \it Fulfillment of  a color confinement criterion due to Kugo-Ojima}
\begin{equation}
 \lim_{Q^2 \rightarrow 0} [Q^2 \Delta_{FP}(Q^2)]^{-1} \equiv 1+u(0) = 0 .
\label{KO}
\end{equation}
However, it should be remarked that these solutions are obtained under the specific truncation of neglecting the two-loop diagrams in the gluon SD equation.  
Nevertheless, recent lattice simulations support some of the above results:
[Bonnet, Bowman, Leinweber and Williams, hep-lat/0002020],
[Langfeld, Reinhardt and Gattnar, hep-lat/0107141],
[Alexandrou, de Forcrand and Follana, hep-lat/0112043],
[Furui and Nakajima, hep-lat/0305010].


\begin{figure}[htbp]
\begin{center}
\includegraphics[height=3cm]{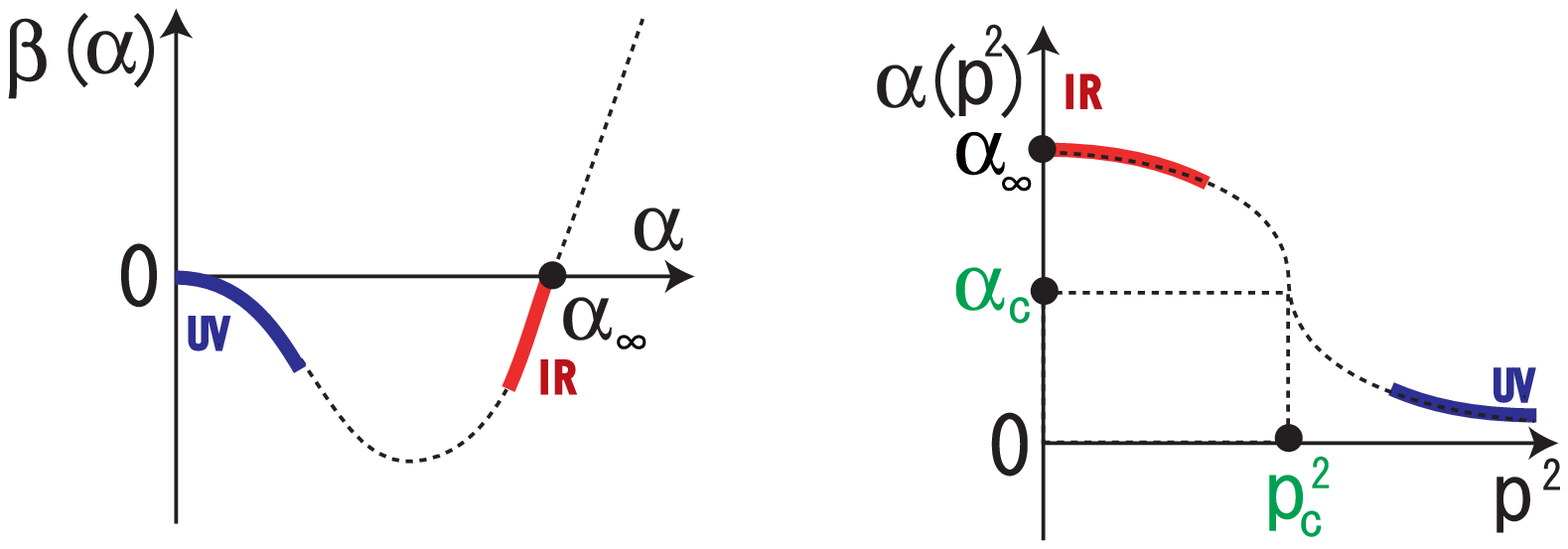}
\includegraphics[height=3cm]{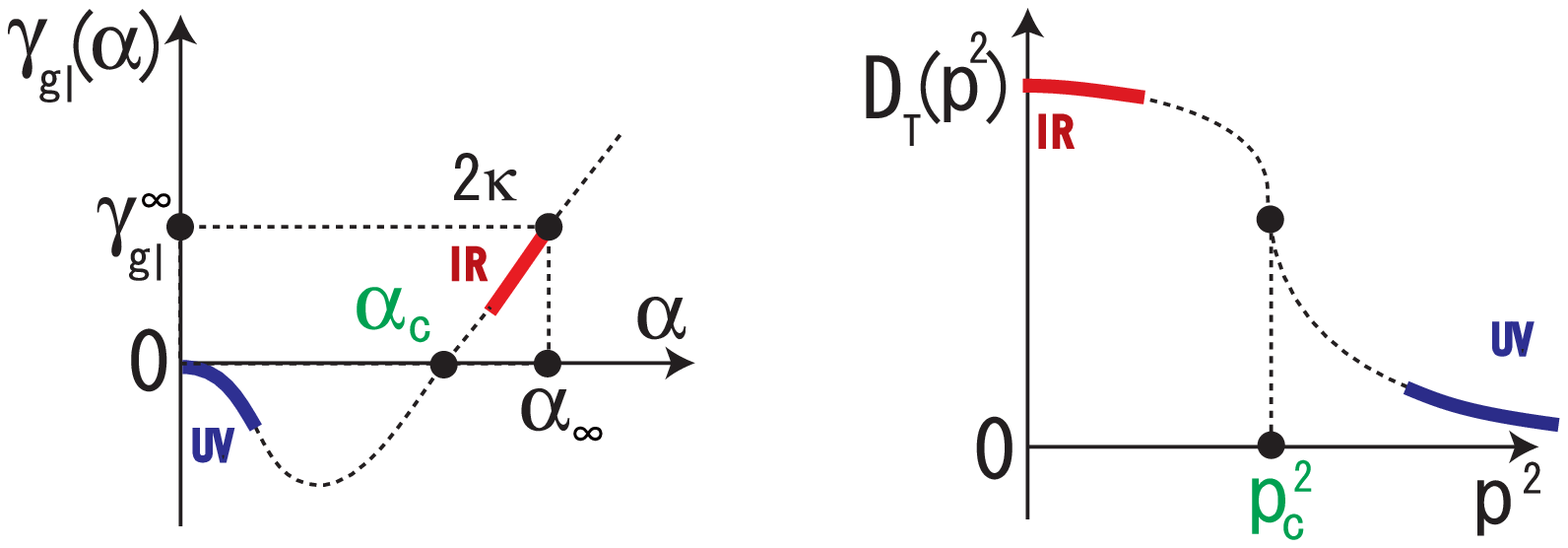}
\includegraphics[height=3cm]{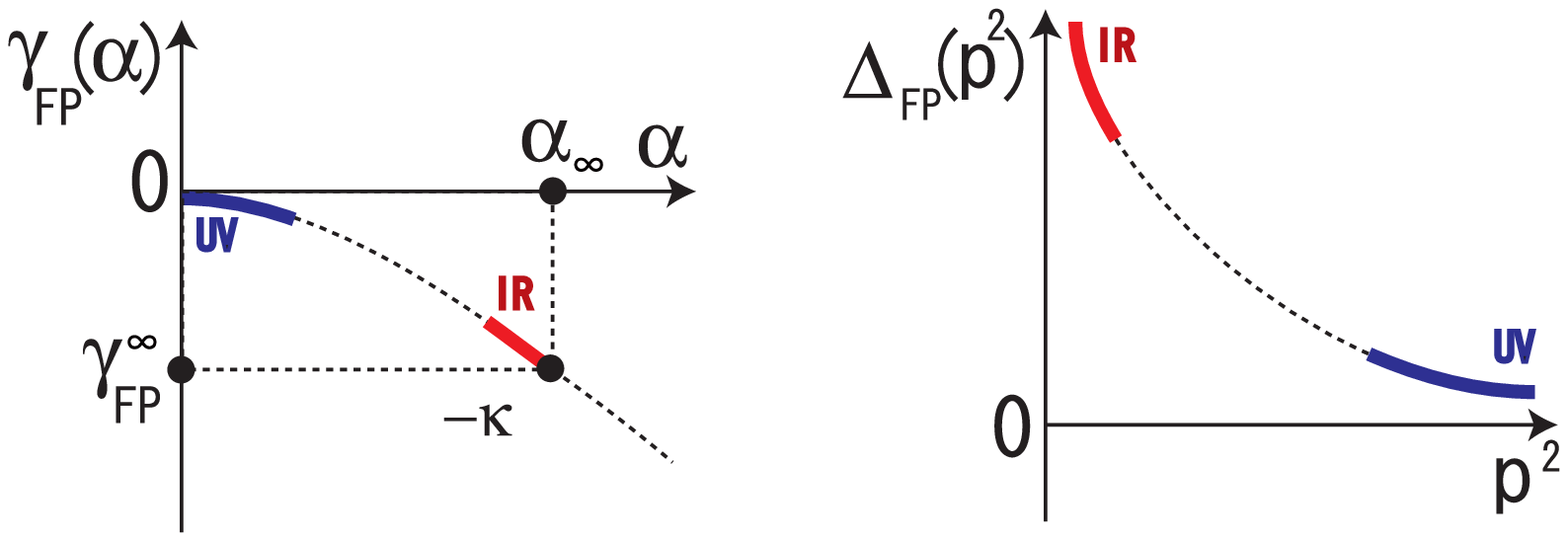}
\caption{\small
The schematic behavior of RG functions, running coupling constant and propagators for gluons and ghosts for Yang--Mills theory in the Lorentz--Landau gauge. 
(a) Beta function vs. Running coupling,
(b) Anomalous dim. of gluon vs. Gluon propagator,
(c) Anomalous dim. of ghost vs. FP Ghost propagator.
}
\label{fig:RGfunct-pfunct2}
\end{center}
\end{figure}


It is obvious that the  above solutions of the SD equations do not correspond to the dynamical masss generation.
However, the latest results of SD equations seem to suggest that 
  {\it $\kappa=0.5$} is allowed as a solution if the two-loop diagrams and hence all the diagrams are included in the SD equations. \cite{Bloch03}
On the other hand, it was argued by the author \cite{Kondo03a} that the axiomatic consideration yields {\it $\kappa=1/2$}: 
\begin{align}
 D_T(Q^2) &\cong \text{const.}+ \mathcal{O}(Q^2)  \downarrow \text{const.} 
\quad \text{as} \quad Q^2 \downarrow 0 ,
\\
 \Delta_{FP}(Q^2) &\cong {1 \over Q^2}(Q^2)^{-1/2} \cong {1 \over (Q^2)^{3/2}} \uparrow \infty  \quad \text{as} \quad Q^2 \downarrow 0 .
\end{align}
These solutions also satisfy (\ref{IRFP}) and (\ref{KO}). 
See Fig.~1. 

 To find a non-perturbative solution with dynamical mass generation is an interesting question in itself.
However, more important issue is to answer the question:
What is the mechanism of generating the gluon mass?
This is because the equation can not be solved without approximation in almost all cases, whereas the mechanism does not depend on the approximation if it is true. 
In fact, the above solutions of the SD equations are 
gauge dependent and 
 not systematically  improvable.

In this talk, we propose the {\it gluon pair condensation} \cite{FK78} as a mechanism of {\it gluon mass generation and color confinement} in $SU(N)$ Yang-Mills theory. 
See Fig.~2. 
A purpose of this work is to construct an approximate {\it solution of QCD which is blessed with both dynamical gluon mass generation and color confinement}. 

\begin{figure}[htbp]
\begin{center}
\includegraphics[height=3cm]{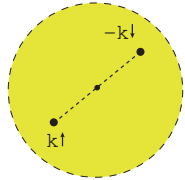}
\caption{\small Gluon pair condensation in  
spin singlet and color singlet channel (center of mass frame). 
}
\label{fig:Pair_condensation}
\end{center}
\end{figure}


For such Bose-Einstein (BE) condensation to take place, we need the net attractive force between gauge boson pair. 
In Yang-Mills theory, gluon self-interactions exist.
See Fig.~3.


\begin{figure}[htbp]
\begin{center}
\includegraphics[height=3cm]{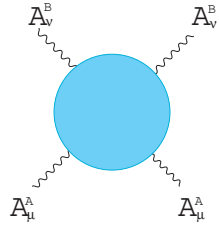}
\caption{\small Interacting pair of gauge bosons. 
}
\label{fig:interacting_bosons}
\end{center}
\end{figure}

In the perturbation theory, short-distance gluon-gluon potential leads to the net attractive force \cite{CS83}, although the details depend on the total spin of the gluon pair.  
See Fig.~4. 
In order to see the long-distance effect realizing the gluon confinement, we must go beyond perturbation theory.


\begin{figure}[htbp]
\begin{center}
\includegraphics[height=3cm]{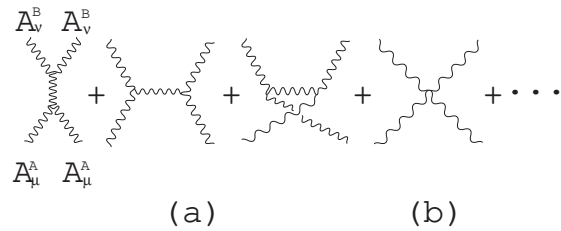}
\caption{\small Perturbative gluon--gluon interaction  to the lowest order (at tree level): (a) attractive interaction for $s-$,$t-$,$u-$channel gluon exchange, (b) attractive or repulsive force depending on the total spin $\bm{S}=\bm{S}_1+\bm{S}_2$ of two-gluons .  
}
\label{fig:perturbative_interaction}
\end{center}
\end{figure}


For this purpose, we set up a new {\it large $N$ expansion} based on a new formulation of Yang-Mills theory.
It is said that 
$N$ is  only one free parameter in $SU(N)$ Yang-Mills theory. 
In general, the large $N$ (or the $1/N$) expansion is
\\
\leftline{$\cdot$ non-perturbative}
\leftline{$\cdot$ systematically  improvable}
\leftline{$\cdot$ gauge symmetry preserving}
\leftline{$\cdot$ related to string theory ['t Hooft,1974]}
We also suggest a new scheme of the coupled SD equation to include all the diagrams. 
For details, see \cite{Kondo03b}.
\section{Yang-Mills theory without the gluon self-interactions}

The Euclidean $SU(N_c)$ Yang-Mills Lagrangian is given by
\begin{align}
\mathcal{L}_{\text{YM}} 
 =   \frac14 {\mathcal F}_{\mu\nu}^2
 =  \frac{1}{4} (\partial_\mu \mathcal{A}_\nu - \partial_\nu \mathcal{A}_\mu + g \mathcal{A}_\mu \times \mathcal{A}_\nu)^2 .
  \label{YM1}
\end{align}
In what follows, we use a notation
$F^2 := F \cdot F $,  
$(F \cdot G) :=   F^A G^A = 2{\rm tr}(FG)$,  
$
  (F \times G)^A := f^{ABC} F^B G^C =-2i {\rm tr}(T^A [F,G])
$,  
and
$
  (F \star G)^A :=  d^{ABC} F^B G^C  = 2 {\rm tr}(T^A \{ F,G \})  
$.

We rewrite the Yang-Mills theory into an equivalent theory in which 
 the gluon self-interactions are eliminated by introducing a number of auxiliary fields 
$\phi, B_{\mu\nu}^A, G_{\mu\nu}, \varphi^A, V_{\mu\nu}^A$
and by choosing the parameters appropriately. 
\begin{align}
  \mathcal{L}_{\text{YM}} 
  =&  {1-\rho^2 \over 4} (\partial_\mu \mathcal{A}_\nu - \partial_\nu \mathcal{A}_\mu)^2 
+{1 \over 4} B_{\mu\nu} \cdot B^{\mu\nu} 
\nonumber\\& 
- {i \over 2} B^{\mu\nu} \cdot {}^*[\rho (\partial_\mu \mathcal{A}_\nu - \partial_\nu \mathcal{A}_\mu) + \sigma g (\mathcal{A}_\mu \times \mathcal{A}_\nu)]
\nonumber\\& 
   + {\sigma_\phi \over 2} \phi^2 + {\sigma_\phi \over 2} \phi (\mathcal{A}_\mu \cdot \mathcal{A}^\mu)
  + {\sigma_G \over 2} G_{\mu\nu} G^{\mu\nu} + {\sigma_G \over 2} G_{\mu\nu} S^{\mu\nu}
  \nonumber\\ &
  + {\sigma_\varphi \over 2} \varphi \cdot \varphi + {\sigma_\varphi \over 2} \varphi \cdot (\mathcal{A}_\mu \star \mathcal{A}^\mu)
  + {\sigma_V \over 2} V_{\mu\nu} \cdot V^{\mu\nu} + {\sigma_V \over 2} V_{\mu\nu} \cdot T^{\mu\nu} ,
  \label{YM4}
\end{align}
where the three-gluon interactions have been eliminated by choosing 
\begin{align}
 \rho = \sigma^{-1} ,
  \label{relation}
\end{align}
and the four-gluon interactions have been eliminated by choosing 
\begin{align}
  \sigma_\phi = {3(\sigma^2-1)g^2 \over N_c}, \
  \sigma_G = {4(1-\sigma^2)g^2 \over N_c}, \
  \sigma_\varphi =  (3/2)(\sigma^2-1)g^2 , \
  \sigma_V =  2(1-\sigma^2)g^2  .
  \label{sigmaparameter}
\end{align}
Here $*$ denotes the Hodge star (duality) operation ${}^*$ which is defined for the second-rank tensor as
$
  {}^*H_{\mu\nu} := \frac{1}{2} \epsilon_{\mu\nu\rho\sigma} H^{\rho\sigma}  
$.

In addition, the GF+FP term is given by
\begin{align}
  \mathcal{L}_{\rm GF+FP} &= \bm{\delta}_{\rm B} \bar{\bm{\delta}}_{\rm B} \left(  {1 \over 2} \mathcal{A}_\mu \cdot \mathcal{A}_\mu  + {\alpha' \over 2} \mathcal{C} \cdot \bar{\mathcal{C}} \right) 
- {\alpha \over 2} \mathcal{B} \cdot \mathcal{B} .
\label{GFglobal2}
\end{align}
By introducing further auxiliary fields $\phi_{FP}, \varphi_{FP}$, the four-ghost self-interactions are eliminated as
\begin{align}
  \mathcal{L}_{\rm GF+FP} 
=&  {1 \over 2\lambda}(\partial_\mu \mathcal{A}_\mu)^2
+  \bar{\mathcal{C}} \cdot \partial_\mu \partial_\mu \mathcal{C}
+ (1-\xi)g    \mathcal{A}_\mu \cdot (\partial_\mu \bar{\mathcal{C}} \times \mathcal{C})
\nonumber\\&
- \xi g  \mathcal{A}_\mu \cdot (\bar{\mathcal{C}} \times \partial_\mu \mathcal{C})
- {1 \over 2}\lambda \xi (1-\xi) g^2 ( \mathcal{C} \times \bar{\mathcal{C}}) \cdot ( \mathcal{C} \times \bar{\mathcal{C}})
\nonumber\\&
 + {\sigma_{FP} \over 2} \phi_{FP}^2 
+ {\sigma_{FP}' \over 2} \varphi_{FP} \cdot \varphi_{FP}   
+ \sigma_{FP} \phi_{FP}  \bar{\mathcal{C}} \cdot \mathcal{C} 
 + \sigma_{FP}' \varphi_{FP} \cdot (\bar{\mathcal{C}} \star \mathcal{C}) ,
 \label{GF+FP3}
\end{align}
where 
\begin{align}
   \sigma_{FP} = \lambda \xi (1-\xi) g^2 , \quad 
 \sigma_{FP}' = \lambda \xi (1-\xi) g^2  ,
\end{align}
and
\begin{equation}
  \lambda :=\alpha+\alpha' , \quad
  \xi :={\alpha'/2 \over \alpha+\alpha'} ={\alpha' \over 2\lambda}.
\end{equation}

Finally, we obtain the total Lagrangin
$
  \mathcal{L}_{\text{YM}}^{\rm tot} = \mathcal{L}_{\text{YM}}+\mathcal{L}_{\rm GF+FP} 
$
 which is at most quadratic in the gluon field $\mathcal{A}_\mu$:
\begin{align}
  \mathcal{L}_{\text{YM}}^{\rm tot} 
  =& 
 {1 \over 4} B_{\mu\nu} \cdot B^{\mu\nu} 
  + {\sigma_\phi \over 2} \phi^2  
  + {\sigma_G \over 2} G_{\mu\nu} G^{\mu\nu}   
  + {\sigma_\varphi \over 2} \varphi \cdot \varphi  
  + {\sigma_V \over 2} V_{\mu\nu} \cdot V^{\mu\nu}  
\nonumber\\& 
+ {\sigma_{FP} \over 2} \phi_{FP}^2 
+ {\sigma_{FP}' \over 2} \varphi_{FP} \cdot \varphi_{FP}   
\nonumber\\& 
  +  \bar{\mathcal{C}} \cdot \partial_\mu \partial^\mu \mathcal{C}
+ \sigma_{FP} \phi_{FP}  \bar{\mathcal{C}} \cdot \mathcal{C} 
 + \sigma_{FP}' \varphi_{FP} \cdot (\bar{\mathcal{C}} \star \mathcal{C})
\nonumber\\& 
+ {1 \over 2} \mathcal{A}_\mu^A \mathcal{K}^{AB}{}^{\mu\nu} \mathcal{A}_\nu^B  + \mathcal{A}_\mu \cdot \mathcal{J}^\mu ,
  \label{YM5}
\end{align}
where we have defined $\mathcal{K}^{AB}_{\mu\nu}$ and $\mathcal{J}_\mu^A$ by  
\begin{align}
 \mathcal{K}^{AB}_{\mu\nu} :=& \delta^{AB} \left[ -(1-\rho^2) (\partial^2 \delta_{\mu\nu} - \partial_\mu \partial_\nu) 
 -  \lambda^{-1} \partial_\mu \partial_\nu  \right]
  \nonumber\\ &
  -i g \sigma f^{ABC} {}^*B_{\mu\nu}^C 
  +  \sigma_\phi \delta^{AB} \delta_{\mu\nu} \phi 
  +   \sigma_\varphi d^{ABC} \delta_{\mu\nu} \varphi^C 
  \nonumber\\ &
  +\sigma_G \delta^{AB} \left( G_{\mu\nu}-{1 \over 4}\delta_{\mu\nu}G_\rho{}^\rho \right)
  +\sigma_V d^{ABC} \left( V_{\mu\nu}^C-{1 \over 4}\delta_{\mu\nu}V^C_\rho{}^\rho \right)   ,
  \nonumber\\ 
  \mathcal{J}_\mu^A :=&  J_\mu^A  - i \rho \partial_\nu {}^*B_{\mu\nu}^{A} + (1-\xi)g   (\partial_\mu \bar{\mathcal{C}} \times \mathcal{C})^A
- \xi g  (\bar{\mathcal{C}} \times \partial_\mu \mathcal{C})^A .
\end{align}
Here  $J_\mu^A$ is the source for the gluon field $\mathcal{A}_\mu^A$. Note that $\mathcal{K}^{AB}_{\mu\nu}$ is at most linear in the auxiliary field and $\mathcal{J}_\mu^A$ contains the bilinear term in the ghost and antighost fields. 
 [The source term for the auxiliary field will be introduced later to avoid the complication.] 

An advantage of this formulation is that we can integrate out the gluon field $\mathcal{A}_\mu$ exactly:
\begin{align}
  S_{\text{EYM}} 
  =& \int_{x} \Big\{
 {1 \over 4} B_{\mu\nu} \cdot B^{\mu\nu} 
  + {\sigma_\phi \over 2} \phi^2  
  + {\sigma_G \over 2} G_{\mu\nu} G^{\mu\nu}   
  + {\sigma_\varphi \over 2} \varphi \cdot \varphi  
  + {\sigma_V \over 2} V_{\mu\nu} \cdot V^{\mu\nu}  
\nonumber\\& 
+ {\sigma_{FP} \over 2} \phi_{FP}^2 
+ {\sigma_{FP}' \over 2} \varphi_{FP} \cdot \varphi_{FP}   
\nonumber\\& 
  +  \bar{\mathcal{C}} \cdot \partial_\mu \partial^\mu \mathcal{C}
+ \sigma_{FP} \phi_{FP}  \bar{\mathcal{C}} \cdot \mathcal{C} 
 + \sigma_{FP}' \varphi_{FP} \cdot (\bar{\mathcal{C}} \star \mathcal{C})
\nonumber\\& 
- {1 \over 2} \mathcal{J}_\mu^A [\mathcal{K}^{AB}{}^{\mu\nu}]^{-1} \mathcal{J}_\nu^B   \Big\} 
+{1 \over 2} \ln {\rm Det} [\mathcal{K}^{AB}{}^{\mu\nu}] .
  \label{YM7}
\end{align}
When $\lambda=0$, $\sigma_{F}=0=\sigma_{FP}'$ and the ghost self-interaction term disappears. 
Therefore, we do not need to introduce $\phi_{FP}$ and $\varphi_{FP}^A$ in (\ref{YM7}) which decouple  from the theory. 

Some remarks are in order. 

1. The resulting theory has one undetermined parameter $\sigma$.  In the classical level, any value of $\sigma$ reproduces the original Yang-Mills theory, as is trivial from the above construction. 
However, in the quantum level, this is not the case.  The renormalization  urges the parameter $\sigma$ to run according to the renormalization scale. 

2. For a special choice $\sigma^2=1$, we have $\rho^2 =\sigma^2=1$ and hence the four coefficients vanish $\sigma_\Phi=0$, which implies that all the auxiliary fields  
$\phi, G_{\mu\nu}, \varphi, V_{\mu\nu}$
except for $B_{\mu\nu}$ 
decouple from the theory. 
Therefore, this case reproduces the field strength formulation \cite{Halpern77} of the Yang-Mills theory, which is identified with the deformation of the topological BF theory \cite{BFYM,KondoII} 
with the Lagrangian:  
\begin{align}
  \mathcal{L}_{\text{BF}} 
  = 
 {1 \over 4} B_{\mu\nu} \cdot B^{\mu\nu} 
- {i \over 2} B^{\mu\nu} \cdot {}^*\mathcal{F}_{\mu\nu}  .
  \label{YM6}
\end{align}
(The signature of $\rho$ and $\sigma$ can be absorbed by the redefinition of the fields $\mathcal{A}_\mu$ and $B_{\mu\nu}$.)
  When $\sigma^2 \not=1$, therefore, we have the scalar field $\phi, \varphi$ and the symmetric tensor fields $G_{\mu\nu}, V_{\mu\nu}$ in addition to the antisymmetric tensor field $B_{\mu\nu}$.

\section{Dynamical gluon mass generation, vacuum condensate and glueball}

If the auxiliary (composite) field $\phi$ develops the non-vanishing vacuum expectation value (VEV), 
\begin{equation}
  \langle \phi \rangle := \phi_0 \not= 0 ,
\end{equation}
the gluons acquire the common mass $M_A$:
\begin{align}
  M_A^2 = g \phi_0 .
\end{align}
We argue that the mechanism of dynamical gluon mass generation is the pair condensation of transverse gluons. 
The VEV is obtained as a non-trivial solution of the gap equation:
\begin{align}
 {1 \over 3(\sigma^2-1)} \phi 
+{1 \over 2} \hat{g}^2 
{\rm tr} (\mathcal{K}^{-1}_{\mu\mu}[\phi])  = 0 ,
\label{gap0}
\end{align}
where $\hat{g}^2$ is the 't Hooft coupling 
$\hat{g}^2:=g^2N_c$ and $\mathcal{K}^{-1}_{\mu\nu}$ is the gluon full propagator in the condensed vacuum. 
We can in principle calculate the effective action $S_E[\phi]$ from which the gap equation is obtained as the saddle-point equation. 
The fluctuation mode $\tilde \phi$ of $S_E[\phi]$ around $\phi_0$ gives the glueball.  For a constant $\phi$, the effective action reduces to the effective potential, i.e., $S_E[\phi]=V(\phi)$. 
The $\phi_0$ gives the absolute minimum of $V(\phi)$. 
See Fig.~5. 

\begin{figure}[htbp]
\begin{center}
\includegraphics[height=3cm]{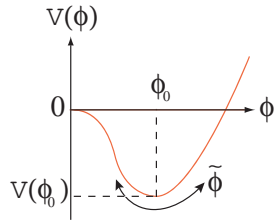}
\caption{The effective potential for the auxiliary field $\phi$. 
}
\label{fig:non-trivial_vev}
\end{center}
\end{figure}


The ratio of the glueball mass $0^{++}$ to the gluon mass is given by 
\begin{align}
  M(0^{++})/M_A = \sqrt{6} \cong 2.45 ,
\end{align}
in the leading order. 

Among a number of the auxiliary fields introduced above, it is the anti-symmetric tensor field $B_{\mu\nu}$ that is most important  from the viewpoint of confinement.
This is because the anti-symmetric tensor field can  couple in a natural way to a surface element of the two-dimensional surface whose boundary is the Wilson loop   
 and this property is desirable for deriving the area law decay of the Wilson loop average, i.e., quark confinement. 
Although the auxiliary fields do not have the kinetic term in the beginning by definition,   the kinetic terms may be generated due to quantum effects.  Then $B_{\mu\nu}$ mediates the confining force. 
For this scenario to be successful, the existence of the condensation $\phi_0$, i.e., non-vanishing vacuum expectation value (VEV) is indispensable, since it provides with the mass scale which is compensated for the dimension introduced through the kinetic term generated. (Note that $\partial_\mu^2/\phi_0$ is dimensionless.)  Without the condensation $\phi_0=0$, $B_{\mu\nu}$ can not acquire the kinetic term.

\section{Vacuum condensate of mass dimension 2}

The existence of {\it vacuum condensate of mass dimension two, i.e., $\langle \mathcal{A}_\mu^2 \rangle \not=0$}, 
was argued to exist recently by several groups.  
The composite operator $\mathcal{A}_\mu^2$ is not gauge invariant, while 
$\bar{\psi}\psi$ and $\mathcal{F}_{\mu\nu}^2$ are gauge invariant. 
Does such a vacuum condensate have any physical meaning?
\\
$\cdot$ The minimum of $ \mathcal{A}_\mu^2$ along the gauge orbit can have a definite physical meaning, as argued by [Zakharov et al.,2001]

\begin{align}
 \delta_\omega \int d^4x \frac{1}{2} \mathcal{A}_\mu^2
= \int d^4x \mathcal{A}_\mu \cdot \delta_\omega \mathcal{A}_\mu
=& \int d^4x \mathcal{A}_\mu \cdot \partial_\mu \omega(x) 
\nonumber\\
=& - \int d^4x \partial_\mu \mathcal{A}_\mu \cdot \omega(x) .
\end{align}
\\
$\cdot$ Lattice simulations + OPE fit performed by 
[Boucaud et al., 2000,2001,2002]
  yield the sizable value 
$\langle \mathcal{A}_\mu^2 \rangle \cong (1.4 \text{GeV})^2$ and the operator product expansion of the running coupling reads 
\begin{align}
  \alpha(p) = \alpha_{pert}(p) \left\{ 1 + R \left( \ln \frac{p}{\Lambda_{QCD}} \right)^{-9/44} 
 \langle \mathcal{A}_\mu \mathcal{A}_\mu  \rangle/p^2   \right\} .
\end{align}
This implies the confining potential at {\it short} distance. (It is argued how the instanton or monopole contributes to this condensate.)

Thus the existence of vacuum condensate with mass dimension two means the existence of non-perturbative power correction even in the high-energy region.

We have proposed a on-shell BRST-invariant composite operator of mass dimension two.  It is given in the partial gauge fixing $G \rightarrow H$ by\cite{Kondo01}:
\begin{align}
  \mathcal{O} := \Omega^{-1} \int d^4x \ {\rm tr}_{G/H} \left[ {1 \over 2} \mathcal{A}_\mu \mathcal{A}^\mu + \lambda i \bar{\mathcal{C}} \mathcal{C} \right] .
\label{mixed}
\end{align}
Therefore,  $\mathcal{O}$ is invariant under the on-shell BRST (and anti-BRST) transformation when 
$
 \xi= {1 \over 2}  \quad \text{or} \quad \lambda=0 .
$
In particular, in the Landau gauge $\lambda=0$, $\partial^\mu  \mathcal{A}_\mu(x)=0$.
Hence $\mathcal{O}$ is  off-shell BRST invariant. 

We have argued in  \cite{Kondo03c} 
\begin{enumerate}
\item[1.] 

The on-shell BRST closed operator $\mathcal{O}$ ($\bm{\delta}_{\rm os}\mathcal{O}=0$) is written as a sum of the gauge-invariant (but nonlocal) operator $\mathcal{O}'$ ($\bm{\delta}_{\omega}\mathcal{O}'=0$) and the on-shell BRST exact part:
\begin{align}
  \mathcal{O} = \mathcal{O}' + \bm{\delta}_{\rm os} \mathcal{O}'' ,
\label{decomp}
\end{align}
where
\begin{align}
  \mathcal{O}' =& \Omega^{-1} 
 \int {d^3 \bm{k} \over (2\pi)^3 2|\bm{k}|} \sum_{\sigma=\pm}  a^A(\bm{k},\sigma) a^A{}^\dagger(\bm{k},\sigma)  ,
\nonumber\\&
\leftarrow \text{gauge-invariant transverse (but nonlocal)}
\nonumber\\
  \mathcal{O}'' =&  \Omega^{-1} 
 \int {d^3 \bm{k} \over (2\pi)^3 2|\bm{k}|} \lambda [i \bar{c}^A(\bm{k}) a^A{}^\dagger(\bm{k},L)+ \text{h.c.}] ,
\nonumber\\&
\leftarrow \text{on-shell BRST exact}
\end{align}
and
\begin{align}
   \bm{\delta}_{\rm os} \mathcal{O}'' 
&=  \Omega^{-1} 
 \int {d^3 \bm{k} \over (2\pi)^3 2|\bm{k}|} [a^A(\bm{k},S) a^A{}^\dagger(\bm{k},L)+ \lambda i \bar{c}^A(\bm{k}) c^A{}^\dagger(\bm{k}) + \text{h.c.}] .
\nonumber\\&
\leftarrow \text{longitudinal and scalar modes cancel the ghost and antighost}
\nonumber
\end{align}

\item[2.] The VEV of $\mathcal{O}$ is equivalent to the VEV of the gauge invariant operator $\mathcal{O}'$ which has zero ghost number and is written in terms of the transverse gauge boson alone, 
\begin{align}
  \langle \mathcal{O} \rangle = \langle \mathcal{O}' \rangle ,
\quad
  \mathcal{O}' = \Omega^{-1} \int d^4x \sum_{\sigma=\pm} {1 \over 2}[A_\mu^{(\sigma)}(x)]^2 .
\end{align}
Therefore, 
$\langle \mathcal{O} \rangle \not= 0$
is a gauge invariant and gauge independent statement.

\item[3.]
 The gauge invariant operator $\mathcal{O}'$ is nonlocal and can be non-linear in the non-Abelian gauge theory.
The transverse gauge boson is both BRST-invariant and gauge invariant, i.e., 
$\bm{\delta}_{\rm B} \mathcal{A}_{\text{phys}}^i(x)=0$, 
$\bm{\delta}_{\rm \omega} \mathcal{A}_{\text{phys}}^i(x)=0$. 

\end{enumerate}



\section{Conclusion and discussion}

We have proposed an equivalent formulation of the Yang-Mills theory in which a sufficient number of auxiliary fields are introduced to eliminate all the gluon self-interactions. 

We have set up a new large $N_c$ expansion of $SU(N_c)$ Yang-Mills theory on a non-perturbative vacuum which is stabilized by transverse  gluon pair condensation. 
Then we have given the Feynman rule of the  large $N_c$ expansion.  
This large $N_c$ expansion is defined on a non-trivial vacuum in which the vacuum condensate of mass dimension two take place. 
A physical meaning of the on-shell BRST invariant composite operator of mass dimension two is discussed. 

We determined the mass ratio of the lowest scalar glueball to the gluon:  $M(0^{++})/M_A = \sqrt{6}\cong2.45$ to the leading order.
This should be compared with the potential model\cite{CS83}
 $M(0^{++})/M_A \cong2.3$. 
For $M_A=600 \sim 700 (MeV)$, $M(0^{++}) = 1.47 \sim 1.72 (GeV)$.

To the leading order, the static potential between a pair of color charges is calculated.  It is given by the sum of the Yukawa-type potential and the linear potential. The non-zero string tension is provided by the condensation.  
The resulting expression is very similar to that \cite{Suzuki88} obtained in the MA gauge assuming the Abelian dominance. 
The string representation is also derived within the same approximation.  The corresponding confining string is nothing but the rigid string due to Polyakov.

\end{document}